\documentstyle[editedvolume,psfig]{crckapb} 
\def\la{\mathrel{\mathchoice {\vcenter{\offinterlineskip\halign{\hfil
 $\displaystyle##$\hfil\cr<\cr\sim\cr}}}
 {\vcenter{\offinterlineskip\halign{\hfil$\textstyle##$\hfil\cr
 <\cr\sim\cr}}}
 {\vcenter{\offinterlineskip\halign{\hfil$\scriptstyle##$\hfil\cr
 <\cr\sim\cr}}}
 {\vcenter{\offinterlineskip\halign{\hfil$\scriptscriptstyle##$\hfil\cr
 <\cr\sim\cr}}}}}
\def\ga{\mathrel{\mathchoice {\vcenter{\offinterlineskip\halign{\hfil
 $\displaystyle##$\hfil\cr>\cr\sim\cr}}}
 {\vcenter{\offinterlineskip\halign{\hfil$\textstyle##$\hfil\cr
 >\cr\sim\cr}}}
 {\vcenter{\offinterlineskip\halign{\hfil$\scriptstyle##$\hfil\cr
 >\cr\sim\cr}}}
 {\vcenter{\offinterlineskip\halign{\hfil$\scriptscriptstyle##$\hfil\cr
 >\cr\sim\cr}}}}}
\def\degr{\hbox{$^\circ$}}
\newcommand{\aj}{AJ}         % Astronomical Journal
\newcommand{\aaa}{A\&A}      % Astronomy and Astrophysics
\newcommand{\aas}{A\&AS}     % Astronomy and Astrophys. Supplement Series
\newcommand{\aar}{A\&AR}     % Astronomy and Astrophysics Review
\newcommand{\apj}{ApJ}       % Astrophysical Journal
\newcommand{\apjl}{ApJL}     % Astrophysical Journal Letters
\newcommand{\mnras}{MNRAS}   % Monthly Notices of the Roy. Astron. Society
\newcommand{\scrm}[1]{\mbox{\scriptsize\rm #1}}

\def\lmean{\mathopen{<}}
\def\rmean{\mathclose{>}}
\newcommand{\etal}{{\it et al.\/}~}

\newcommand{\ie}{{\it i.e.},\ }
\newcommand{\eg}{{\it e.g.},\ }

\newcommand{\Mzon}{M$_{\odot}$}

\newcommand{\MdI}{$M_{\scrm{d,\,{\sc iras}}}$}
\newcommand{\MdO}{$M_{\scrm{d,\,opt}}$}

\newcommand{\kms}{\mbox{km s$^{-1}$}}

\newcommand{\Mpc}{Mpc$^{-1}$} 
\newcommand{\Ha}{H$\alpha$}

\newcommand{\NII}{[{\sc N$\,$ii}]}
\newcommand{\HI}{{\sc H$\,$i}}

\newcommand{\um}{\mbox{\rm $\mu$m}}

\begin{opening}
\title{THE DISTRIBUTION OF DUST AND GAS IN ELLIPTICAL GALAXIES}%\protect\\
\author{PAUL GOUDFROOIJ}
\institute{European Southern Observatory \\ Karl-Schwarzschild-Strasse 2,
   D-85748 Garching, Germany}
\end{opening}

\runningtitle{THE DISTRIBUTION OF DUST AND GAS IN ELLIPTICAL GALAXIES}

\begin{document}

% The \begin{document} command comes after the \end{opening}
% command.

\begin{abstract}
\baselineskip=0.90\normalbaselineskip 
{\small 
Results from {\sl IRAS\/} and recent optical CCD surveys are examined to
discuss the distribution and origin of dust and ionized gas in elliptical
galaxies. In strong contrast with the situation among spiral galaxies, masses
of dust in elliptical galaxies as derived from optical extinction are an
order of magnitude {\it lower\/} than those derived from {\sl IRAS\/}
data. I find that this dilemma can be resolved by a 
{\it diffusely distributed component\/} of dust which is not detectable in
optical data. \\
\mbox{\small ~~~~~The} morphology of dust lanes and their association with
ionized gas in elliptical galaxies argues for an external origin of both
components of the ISM.   
}
\end{abstract}

\section{Introduction: Dust and Gas in Elliptical Galaxies}

Recent advances in instrumental sensitivity have challenged the very
definition of elliptical galaxies in Hubble's galaxy classification scheme.
It is now clear that ellipticals contain a complex, diverse ISM, primarily in
the form of hot ($T \sim 10^7$\,K), X-ray-emitting gas, with masses $\la 
10^{11}$ \Mzon. Small amounts of \HI, H$_2$, ionized gas, and dust have been
detected as well in many ellipticals (\eg Bregman \etal \citeyear{breg+92}).

Unlike the situation in spiral galaxies, physical and evolutionary
relationships between the various components of the ISM in ellipticals
are not yet understood. A number of theoretical concepts have been
developed for the secular evolution of the different components of the
ISM of ellipticals. 
The two currently most popular concepts are {\it (i)\/} the ``cooling flow'' 
picture in which mass loss from stars within the galaxy, heated to $10^6 -
10^7$ K by supernova explosions and collisions between expanding stellar
envelopes during the violent galaxy formation stage, quiescently cools and
condenses (cf.\ the review of Fabian \etal \citeyear{fabi+91}) and
{\it (ii)\/} the ``evaporation flow'' picture in which clouds of dust and gas
have been  accreted during post-collapse galaxy interactions. Subsequent
heating (and  evaporation) of the accreted gas is provided by thermal
conduction in the hot,  X-ray-emitting gas and/or star formation (cf.\ de Jong
\etal \citeyear{dej+90}; Sparks \etal \citeyear{spa+89}).  

The first direct evidence for the common presence of cool ISM in
ellipticals was presented by Jura \etal (\citeyear{jura+87}) who used {\sl
IRAS\/} ADDSCANs and found that $\ga 50$\% of nearby, bright ellipticals were
detected at 60 and 100 \um. Implied dust masses were of order $\sim 10^4 -
10^6$ \Mzon\ (using H$_0$ = 50 \kms\ \Mpc). 
Interestingly, there are several X-ray-emitting ellipticals with suspected
cooling flows (cf.\ Forman \etal \citeyear{fjt85}) among the {\sl
IRAS\/} detections. The presence of dust in such 
objects is surprising, since the lifetime of a dust grain against collisions
with hot ions (``sputtering'') in hot gas with typical pressures 
$nT\sim10^5$\,cm$^{-3}$\,K is only $10^6 - 10^7$ yr (Draine \&\ Salpeter
\citeyear{drasal79}). What is the origin of this dust, and how is it
distributed?  

In order to systematically study the origin and fate of
the ISM of elliptical galaxies, we have recently conducted a deep, systematic
optical survey of a complete, blue magnitude-limited sample of
56 elliptical galaxies drawn exclusively from the RSA catalog (Sandage
\&\ Tammann \citeyear{rsa81}).  
Deep CCD imaging has been performed through both broad-band filters
and narrow-band filters isolating the nebular \Ha+\NII\ emission lines.

In this paper I combine results from this survey with the {\sl IRAS\/} data to
discuss the distribution and origin of dust and gas in ellipticals. Part of
this paper is based on %findings reported in 
Goudfrooij \& de Jong (\citeyear{paper4}, hereafter Paper IV). 

\section{Detection Rates of Dust from Optical Surveys}

Optical observations are essential for establishing the presence and
distribution of dust and gas in ellipticals, thanks to their high spatial
resolution. A commonly used optical technique to detect dust is by inspecting 
color-index (\eg $B-I$) images in which dust shows up as distinct, reddened
structures with a morphology different from the smooth underlying distribution
of stellar light (\eg Goudfrooij \etal \citeyear{paper2}, hereafter 
Paper II). However, a strong limitation of optical
detection methods (compared to the use of {\sl IRAS\/} data) is that only dust
distributions that are sufficiently 
different from that of the stellar light (\ie dust lanes, rings, or patches)
can be detected. Moreover, detections are limited to nearly edge-on dust
distributions (\eg no dust lanes with inclinations $\ga 35$\degr\ have been
detected, cf.\ Sadler \& Gerhard \citeyear{sg85}; Paper II). Thus, the optical
detection rate of dust (currently 41\%, cf.\ Paper II) represents a firm lower
limit. Since an inclination of 35\degr\ is equivalent to about half the total
solid angle on the sky, one can expect the {\it true\/} detection rate to be
about twice the measured one (at a given detection limit for dust
absorption), which means that {\it the vast majority of ellipticals could
harbor dust lanes and/or patches}. 

\section{The Dust/Ionized Gas Association: Clues to their Origin}

Optical emission-line surveys of luminous, X-ray-emitting ellipticals have
revealed that these galaxies often contain extended regions of ionized gas
(\eg Trinchieri \& di Serego Alighieri \citeyear{tdsa91}), which have been
argued to arise as thermally instable regions in a ``cooling flow''. As
mentioned before, the  emission-line regions in these galaxies are suspected
to be dust-free in view of the very short lifetime of dust grains.  However,
an important result of our optical survey of ellipticals is the finding that
emission-line regions are essentially {\it always\/} associated with
substantial dust absorption 
%
% DEAR DAVID BLOCK, THIS IS WHERE TO PUT A REFERENCE TO MY COLOR PICTURE 
% IN CASE IT IS SELECTED FOR DISPLAY IN THE PROCEEDINGS. I PUT THE TEXT TO BE
% INCLUDED BELOW, COMMENTED BY A PERCENT SIGN. IF THE COLOR PICTURE WILL
% INDEED BE SELECTED, PLEASE COMMENT OUT THE FIRST LINE AFTER THE LINE THAT
% YOU WOULD THEN UNCOMMENT. 
% 
% (see Plate XX; Paper II; see also Macchetto \& Sparks \citeyear{macspa91}), 
(Paper II; see also Macchetto \& Sparks \citeyear{macspa91}), 
which is difficult to account for in the
``cooling flow'' scenario. This dilemma can however  be resolved in the
``evaporation flow'' scenario (de Jong \etal \citeyear{dej+90}) in which the
ISM has been accreted from a companion galaxy. 

Closely related to the origin of the dust and gas is their dynamical state,
\ie whether or not their motions are already settled in the galaxy potential. 
This question is, in turn, linked to the intrinsic shape of ellipticals,
since in case of a settled dust lane, its morphology indicates 
a plane in the galaxy in which stable closed orbits are allowed (\eg Merritt
\& de Zeeuw \citeyear{merdez83}). These issues can be studied best 
in the inner regions of ellipticals, in view of the short relaxation 
time scales involved, allowing a direct relation to the intrinsic shape of the
parent galaxy. 
A recent analysis of properties of {\it nuclear\/} dust in 64 ellipticals
imaged with HST has shown that dust lanes are
randomly oriented with respect to the apparent major axis of the galaxy (van
Dokkum \& Franx \citeyear{vdf95}). Moreover, the dust lane is significantly
misaligned with the {\it kinematic\/} axis of the stars for almost all
galaxies in their sample for which stellar kinematics are available. This
means that {\it even at these small scales}, the dust and stars are generally
dynamically decoupled, which argues for an external origin of the dust. This
conclusion is strengthened by the decoupled kinematics of stars and gas in
ellipticals with {\it large-scale\/} dust lanes (\eg Bertola \etal
\citeyear{bert+88}).  

\section{The ``Dust Mass Discrepancy'' among Elliptical Galaxies} 

As mentioned in the introduction, dust in ellipticals has been detected by 
optical as well as far-IR surveys. Since the optical and far-IR
surveys yielded quite similar detection rates, one is tempted to conclude that
both methods trace the same component of dust. In this Section, this point
will be addressed by discussing the distribution of dust in ellipticals. 

The methods used for deriving dust masses from optical extinction
values and from the IRAS flux densities at 60 and 100 $\mu$m, and the
limitations involved in these methods, are detailed upon in Goudfrooij \etal
(\citeyear{paper3}, hereafter Paper III) and Paper IV. 
It is found that the dust masses estimated from the optical extinction are
significantly {\it lower\/} than those estimated from the far-IR
emission (see Paper IV). Quantitatively, 
the average ratio $\lmean M_{\scrm{d,\,{\sc iras}}}/M_{\scrm{d,\,opt}}
\rmean$ = $8.4 \pm 1.3$ for the galaxies in our ``RSA sample'' for which the
presence of dust is revealed by both far-IR emission and optical
dust lanes or patches. 

I should like to emphasize that this ``dust mass discrepancy'' among
ellipticals is quite remarkable, since the situation is {\it significantly
different\/} in the case of spiral galaxies: Careful analyses of deep
multi-color imagery of dust extinction in spiral galaxies (\eg Block \etal
\citeyear{bloc+94}; Emsellem \citeyear{eric95})
also reveal a discrepancy between dust masses derived from optical and {\sl
IRAS\/} data, {\it but in the other sense, i.e.,\/} $M_{\scrm{d,\,{\sc iras}}}
\ll M_{\scrm{d,\,opt}}${\it \,!}~  This can be understood since the {\sl
IRAS\/} measurements were sensitive to ``cool'' dust with temperatures
$T_{\scrm{d}} \ga 25$\,K, but much less to ``cold'' dust at lower temperatures
which radiates predominantly at wavelengths beyond 100 $\mu$m (\eg Young \etal
\citeyear{young+86}). Since dust temperatures of order 20 K and lower are
appropriate to spiral galaxies (Greenberg \& Li \citeyear{greeli95} and
references therein), dust masses derived from the {\sl IRAS\/} 
data are strict {\it lower limits\/} by nature. 
Evidently, the bulk of the dust in spiral disks is too cold to emit
significantly at 60 and 100 $\mu$m, but still causes significant extinction
of optical light.  
Interestingly, $T_{\scrm{d}} \la 20$\,K is {\it also\/}
appropriate to the outer parts of ellipticals (cf.\ Paper IV),
underlining the significance of the apparent ``dust mass discrepancy'' among
ellipticals. What could be the cause\,?

\noindent {\it Orientation effects\,?\/}~ 
If the discrepancy would be due to an orientation effect, the ratio \MdI/\MdO\
would be inversely proportional to cos\,$i$,
where $i$ is the inclination of the dust lane with respect to the line of
sight.  However, we have measured inclinations of regular, uniform
dust lanes in ellipticals from images shown in homogeneous optical
CCD surveys, and found
that the relation between \MdI/\MdO\ and cos\,$i$  is a scatter plot (cf.\
Fig.\ 1 of Paper IV). Thus, the effect of orientation on the dust mass
discrepancy must be weak if present at all.  This suggests that the dust in
the lanes is concentrated in dense clumps with  a low volume filling factor.  

\begin{figure}%[t]
%\vspace*{4.8cm}
\centerline{\psfig{figure=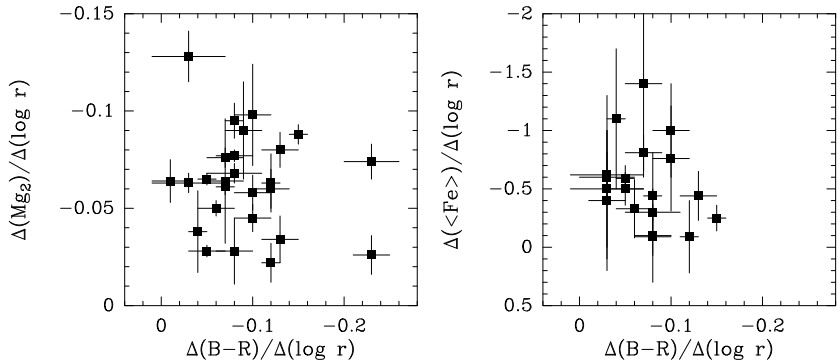,angle=-90.,height=4.8cm}}%
\caption[]{The relation of radial $B-R$ color gradients with radial gradients
of the stellar line-strength indices Mg$_2$ {\sl (left panel)\/} and $\lmean
\mbox{Fe} \rmean$ {\sl (right panel)} for all ellipticals for which
both pairs of quantities have been measured to date. Data taken from Peletier
(\citeyear{pele89}), Carollo \etal (\citeyear{caro+93}), Davies \etal
(\citeyear{dsp93}), Carollo \& Danziger (\citeyear{cardan94}), and Paper I.  
}
\label{fig:gradients}
\end{figure}

\noindent {\it Diffusely distributed dust\,?\/}~
Having eliminated the effect of orientation, the most plausible way
out of the dilemma of the dust mass discrepancy is to
postulate an additional, diffusely distributed component of dust, which is  
therefore virtually undetectable by optical methods. We note that this diffuse
component of dust is not unexpected: the late-type stellar population of
typical giant ellipticals  ($L_B = 10^{10} - 10^{11}$ L$_{\odot}$) has
a substantial present-day mass loss rate ($\sim$ 0.1 $-$ 1 \Mzon\
yr$^{-1}$ of gas and dust; cf.\ Faber \&\ Gallagher \citeyear{fg76}) which can
be expected to be diffusely distributed.  
An interesting potential way to trace this diffuse component of dust is
provided by radial color gradients in ellipticals. With very 
few significant exceptions, ``normal'' ellipticals show a global
reddening toward their centers, in a sense approximately linear with
log\,(radius) (Goudfrooij \etal \citeyear{paper1} (hereafter Paper I) and
references therein). This is usually interpreted as 
gradients in stellar metallicity, as metallic line-strength indices show a
similar radial gradient (\eg Davies \etal \citeyear{dsp93}). However, 
compiling all measurements published to date on color- and line-strength
gradients within ellipticals shows no obvious correlation (cf.\ Fig.\ 
\ref{fig:gradients}), suggesting that an additional process is (partly)
responsible for the color gradients. 
Although the presence of dust in ellipticals is now beyond dispute, the
implications of dust extinction have been generally discarded in the
interpretation of color gradients. However, recent Monte Carlo simulations of
radiation transfer within ellipticals by Witt \etal (\citeyear{wtc92},
hereafter WTC) 
and Wise \& Silva \shortcite{wisesil96} 
have demonstrated that a diffuse distribution of dust
throughout ellipticals can cause significant color gradients even with modest
dust optical depths. 

\begin{figure}%[t] 
%\vspace*{7.5cm}
\centerline{ 
\psfig{figure=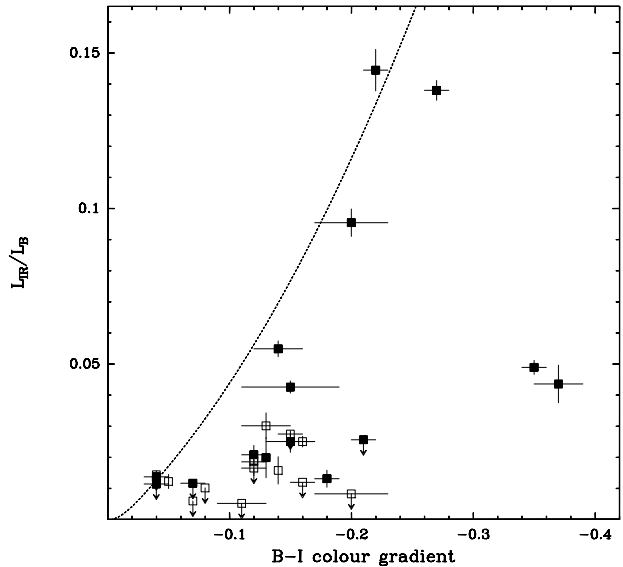,bbllx=70pt,bblly=16pt,bburx=545pt,bbury=538pt,angle=-90.,height=7.5cm}}%height=12.cm}} 
\caption[]{The relation of $L_{IR}$/$L_B$ 
with radial $B-I$ color gradients (defined as 
$\Delta\,$($B-I$)/$\Delta\,$(log r)) for elliptical galaxies 
in the ``RSA sample'' (cf.\ Paper I). Filled squares represent galaxies
detected by IRAS showing optical evidence for dust, and open squares
represent  galaxies detected by IRAS without optical evidence for
dust.  Arrows pointing downwards indicate upper limits to 
$L_{IR}$/$L_B$. 
The dotted line represents the color gradient expected from
differential extinction by a diffuse distribution of dust (see text). Figure
taken from Paper IV.}
\label{IV:lirlbdbidr}
\end{figure}

We have used WTC's ``Elliptical'' model to
predict dust-induced color gradients appropriate to the far-IR
properties of ellipticals in the ``RSA sample'', as derived from the {\sl
IRAS\/} data.  The model features King \shortcite{king62} profiles for the
radial density distributions of both stars and dust:  
\[
\rho(r) = \rho_0\, \left[1 +
\left(\frac{r}{r_0}\right)^2\right]^{-\alpha/2} 
\]
where $\rho_0$ represents the central density and $r_0$ is the core radius.
The ``steepness'' parameter $\alpha$ was set to 3 for the stars, and to 1 for
the dust [The reason for the low value of $\alpha$ for the dust distribution
is that it generates color gradients that are linear with log\,($r$), as
observed (Paper I; Wise \& Silva \citeyear{wisesil96})]. Using this model,
far-infrared-to-blue  luminosity ratios and color gradients 
have been derived as a function of the total optical depth $\tau$
of the dust (\ie the total dust mass). 
The result is plotted in Fig.\ \ref{IV:lirlbdbidr}. 
It is obvious that color gradients in elliptical galaxies are generally
larger than can be generated by a diffuse distribution of dust throughout the
galaxies according to the model of WTC. This is as expected, since color
gradients should be partly due to stellar population gradients as well.
However, {\it none\/} of the galaxies in this sample has a color gradient
significantly {\it smaller\/} than that indicated by the model of WTC. I argue
that this is caused by a ``bottom-layer'' color gradient due to differential
extinction, which should be taken seriously in the interpretation of color
gradients in ellipticals. 

We have checked whether the assumption of the presence of a diffusely
distributed component is also {\it energetically\/} consistent with the
available {\sl IRAS\/} data. To this end, we computed heating rates for dust
grains as a function of galactocentric radius. We assumed heating by {\it
(i)\/} stellar photons, using the radial surface brightness profiles from
Paper I, and {\it (ii)\/} hot electrons in X-ray-emitting gas, if appropriate.
Radial dust temperature profiles are derived by equating the heating rates
to the cooling rate of a dust grain by far-IR emission. 

Using the derived radial dust temperature profiles, we reconstructed {\sl
IRAS\/} flux densities for both the optically visible component and the
(postulated) diffusely distributed component. Apparently regular dust lanes
were assumed to by circular disks of uniform density, reaching down to the
galaxy nucleus.  After subtracting the contribution of the optically visible
component of dust to the {\sl IRAS\/} flux densities, the resulting flux
densities were assigned to the diffuse component. Using the WTC model
calculations (cf.\ Fig.\ \ref{IV:lirlbdbidr}), $L_{IR}$/$L_B$ ratios were
translated into total optical depths of the dust (and hence dust mass column
densities). Dividing the dust masses of the diffusely distributed component by
the dust mass column densities, outer galactocentric radii for the diffusely
distributed dust component were derived (typical values were $\sim 2$ kpc).
Finally, the {\sl IRAS\/} flux densities were constructed from the masses of
the diffusely distributed component by integrating over spheres (we refer the
reader to Paper IV for details).  

\begin{figure}[t]
%\vspace*{6.5cm}
\centerline{ 
\psfig{figure=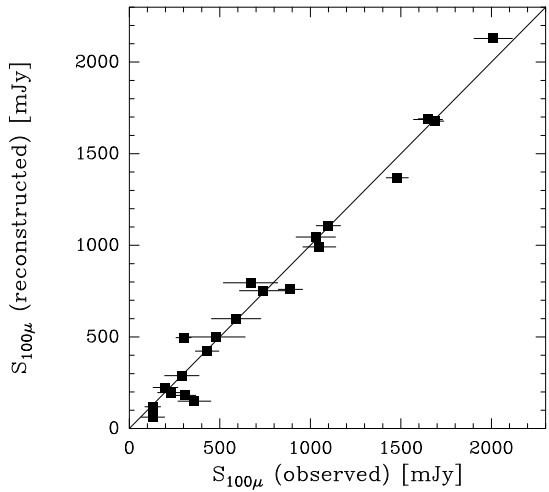,bbllx=23pt,bblly=71pt,bburx=466pt,bbury=467pt,angle=0.,height=6.5cm}}%height=12.cm}} 
\caption[]{The 100 \um\ flux densities reconstructed from our calculations
of the dust temperatures in elliptical galaxies (see text) versus the observed
100 \um\ flux densities (and their 1\,$\sigma$ 
error bars). The solid line connects loci with ``reconstructed = observed''. }
\label{fig:compare100}
\end{figure}

A comparison of the observed and reconstructed {\sl IRAS\/} flux
densities (cf.\ Fig.\ \ref{fig:compare100}) reveals that the {\it
observed\/} {\sl IRAS\/} flux densities can {\it in virtually all
elliptical galaxies in the RSA sample\/} be reproduced {\it within the
1$\,\sigma$ uncertainties\/} by assuming two components of dust in
elliptical galaxies:\ an optically visible component in the form of
dust lanes and/or patches, and a newly postulated dust component which
is diffusely distributed within the inner few kpc from the center of
the galaxies.  

We remind the reader that we have only considered dust which was detected by
{\sl IRAS}, \ie with $T_{\scrm{d}} \ga 25$\,K. 
In reality, the postulated diffuse component of dust in elliptical
galaxies may generally be expected to extend out to where the dust temperature
is lower. Observations with the {\sl Infrared Space Observatory
(ISO)\/} of the RSA sample of elliptical galaxies are foreseen, and
may reveal this cooler dust component in ellipticals.

\paragraph{Acknowledgments.} I am very grateful to the SOC for allowing me to
participate in this great conference. It is also a pleasure to thank Drs.\
Teije de Jong, Leif Hansen, Henning J{\o}rgensen, and Hans-Ulrik
N{\o}rgaard-Nielsen for their various contributions to this project.

\end{document}